%% file: CliffordRBandNISTRB.tex
\documentclass[aps,pra,english,10pt,a4,nofootinbib,twocolumn,notitlepage,superscriptaddress]{revtex4-1}
\include{macros}

\usepackage{mdframed}
\usepackage{graphicx}
\usepackage[toc,page]{appendix}
\usepackage{color}
\usepackage{amsmath, amssymb,amsthm}
\usepackage{mathtools}
\usepackage{multirow}
\usepackage{hyperref}
\usepackage{algorithm}
\usepackage{enumitem}
\usepackage{algpseudocode}
\usepackage{graphicx}
\usepackage{cancel}
\usepackage{verbatim}
\usepackage{array}
\newcolumntype{P}[1]{>{\centering\arraybackslash}p{#1}}
\usepackage{cleveref}
\begin{document}
	\title{Randomized Benchmarking under Different Gatesets}
	\author{Kristine Boone}
	\affiliation{Institute for Quantum Computing and the Department of Applied
		Mathematics, University of Waterloo, Waterloo, Ontario N2L 3G1, Canada}
	\affiliation{Quantum Benchmark Inc., 51 Breithaupt, Kitchener, Ontario, Canada N2H5G5}
	\author{Arnaud Carignan-Dugas}
	\affiliation{Institute for Quantum Computing and the Department of Applied
		Mathematics, University of Waterloo, Waterloo, Ontario N2L 3G1, Canada}
	\affiliation{Quantum Benchmark Inc., 51 Breithaupt, Kitchener, Ontario, Canada N2H5G5}
	\author{Joel J. Wallman}
	\affiliation{Institute for Quantum Computing and the Department of Applied
		Mathematics, University of Waterloo, Waterloo, Ontario N2L 3G1, Canada}
	\affiliation{Quantum Benchmark Inc., 51 Breithaupt, Kitchener, Ontario, Canada N2H5G5}
	\author{Joseph Emerson}
	\affiliation{Institute for Quantum Computing and the Department of Applied
		Mathematics, University of Waterloo, Waterloo, Ontario N2L 3G1, Canada}
	\affiliation{Quantum Benchmark Inc., 51 Breithaupt, Kitchener, Ontario, Canada N2H5G5}
	\affiliation{Canadian Institute for Advanced Research, Toronto, Ontario M5G 1Z8, Canada}

\begin{abstract}
	We provide a comprehensive analysis of the differences between two important standards for randomized benchmarking (RB): the Clifford-group RB protocol proposed originally  in Emerson et al (2005) and Dankert et al (2006), and a variant of that RB protocol proposed later by the NIST group in Knill et al, PRA (2008). While these two protocols are frequently conflated or presumed equivalent, we prove that they  produce distinct exponential fidelity decays leading to differences of up to a factor of 3 in the estimated error rates under experimentally realistic conditions. These differences arise because the NIST RB protocol does not satisfy the unitary two-design condition for the twirl in the Clifford-group protocol and thus the decay rate depends on non-invariant features of the error model.  Our analysis provides an important first step towards developing definitive standards for benchmarking quantum gates and a more rigorous theoretical underpinning for the NIST protocol and other RB protocols lacking a group-structure. We conclude by discussing the potential impact of these differences for estimating fault-tolerant overheads.  \end{abstract}

\maketitle

\section{Introduction}

Clifford-group randomized benchmarking (RB)
\cite{Emerson2005,dankert2006exact} has become the \emph{de facto}
standard tool for assessing and optimizing the quantum control
required for quantum computing systems by estimating error rates
associated with sets of elementary gates operations. It has
been known for some time that this protocol leads to an invariant
exponential decay \cite{Emerson2005,dankert2006exact,Levi2007}
because it is equivalent to a  sequence of twirls  \cite{Levi2007} with
unitary-two designs \cite{dankert2006exact}.  
More recently, the robustness of the Clifford-group
RB protocol has been supported by a
rigorous theoretical framework, including proofs
that an exponential fidelity decay will be observed under very broad
experimental conditions, including essentially arbitrary state
preparation and measurement errors \cite{Magesan2011, Magesan2012a} and
gate-dependent errors \cite{Wallman2018GD}, as well as proofs that the 
observed error rate relates directly to a well-defined notion of 
gate-fidelity \cite{Wallman2018GD, Proctor2017,carignan2018randomized},  
which fully overcome recent  concerns about relating measured RB error rates 
to a meaningful concept of gate-fidelity under gate-dependent errors \cite{Proctor2017}.

While a wide-variety of group-based generalizations of RB have been 
proposed in recent years, \textit{e.g.}~\cite{barends2014rolling, 
Dugas2015, Cross2016, Wallman2016a, Wallman2015b, Wallman2018GD}, 
in this Letter we focus on  clarifying the physical relevance of a standing conflation in the literature 
between the now standard  Clifford-group RB protocol proposed in 
\cite{Emerson2005,dankert2006exact} and an alternate version of RB  
proposed later by NIST \cite{Knill2008}.  As described below, these are 
distinct protocols that measure distinct properties of the error model 
and thus can produce different error rate estimates under the same, realistic experimental conditions. Moreover, because the NIST 
protocol does not admit a closed-group or unitary two-design structure, the rigorous theoretical framework  justifying 
Clifford-group RB does not trivially extend to support the physical 
interpretation and robustness of NIST RB.

In this Letter we identify the operationally-relevant differences between 
the Clifford-group RB protocol and the NIST version of RB 
which clarifies how they can lead to very different error rate estimates 
given the same error model (as defined in terms of the elementary control pulses).  
We then provide the first rigorous proof that the NIST 
RB protocol does indeed produce an exponential decay under 
gate-independent error models. This is an important step toward  
developing a theoretical justification for the NIST protocol and other 
RB protocols that do not admit a group-structure in the case of gate-dependent errors and the ultimate goal 
of a theoretical framework within which error reconstruction under RB 
protocols with different gate sets can be extracted in a unified and 
consistent manner.   Our analysis is thus also essential for comparing 
cross-platform benchmarking methods and standards for quantum 
computing.  As a second contribution, we numerically explore the size 
and scope of the quantitive differences in estimated error rates that 
arise under each  of the protocols for a variety of physically relevant 
error models and pulse-decompositions, and observe that experimentally 
estimated error rates can differ by as much as a factor of 3 in typical cases. 
We conclude by discussing how these differences are relevant for 
detecting gate-dependent 
errors, and estimating fault-tolerant overheads under quantum error 
correction.

\vspace{-1em}
\section{Background and Motivation}

The original proposal for randomized benchmarking from Emerson \textit{et al.~}\cite{Emerson2005} considered 
implementing long sequences of quantum gates drawn \textit{uniformly} at 
random from the group SU($d$) for any quantum systems with Hilbert space dimension $d$.  That work proved that  
the measured fidelity would follow an exponential decay with a decay 
rate that is fixed uniquely by the error 
model, that is, the measured decay rate would not depend on the 
choice of initial state or the specific random quantum gate sequences.

\begin{minipage}[t!]{0.95 \linewidth}
	\raggedright
	\begin{protocol}[Standard Clifford-group RB, as described in \cite{Emerson2005,dankert2006exact}]
		\hrule
		\vspace{1mm}
		~\label{proto:cliff_rb}
		\vspace{1mm}
		\begin{enumerate}[label=\textbf{\arabic*}.]
			\hrule
			\item Sample a set of $m$ gates $G_i$ picked independently and uniformly at random from the Clifford group ${\bf C}$ defined in \cref{eq:clifford_gateset};
			\item Determine the recovery gate $ G_{m+1}$ (see text below);
			\item Prepare a state $\rho \approx |0 \rangle \langle 0 |$; 
			\item Perform the sampled gates from step $1$, followed by the recovery gate $G_{m+1}$ determined in step 2:\\
			$\tilde{G}_{m+1:1} = \tilde{G}_{m+1}\circ\ldots\circ \tilde{G}_1$;
			\item 
			Measure a POVM $\{Q,\mbb I -Q\}$, where the first observable is $Q\approx G_{m+1:1}(|0 \rangle \langle 0 |)$, and respective 
			outcome labels are $\{\text{``recovery'', ``non-recovery''}\}$;
			\item Repeat steps 3--5 a number times to estimate the probability
			of observing the ``recovery'' event ${\rm Pr}(\text{``recovery''}|\{G_i\},m)=\tr Q \tilde{G}_{m+1:1}(\rho)$;
			\item Repeat steps 1--6 for $s$ different sets of $m$ randomly sampled gates $\{G_i\}$; 
			\item Repeat steps for 1--7 for different values of $m$ of random gates. 
			\item Fit the estimated recovery probabilities to the decay model
			\begin{align}
			A_{\bf C}~ p_{\bf C}^m + B_{\bf C}~; \label{eq:cliff_decay_model}
			\end{align}
			\item Estimate the Clifford gate-set infidelity through 
			\begin{align}
			r_{\bf C}= (1-p_{\bf C})/2~. \label{eq:cliff_infidelity}
			\end{align}
			\hrule
		\end{enumerate}
	\end{protocol}
	\vspace{0.1mm}
\end{minipage}

This protocol suffered from two limitations: the random gates were drawn from a continuous set, which is impractical even for $d=2$, 
and the protocol would not be efficient for large systems because a typical random element of SU($d$) 
requires exponentially long gate sequences under increasing numbers of qubits. Additionally, in that limit the inversion gate may not be computed efficiently.

However, practical and efficient solutions to both of these problems  were proposed in 
Dankert \textit{et al.}~\cite{dankert2006exact} in 2006, which proved and observed that 
 drawing gates uniformly at random from the Clifford group would 
lead to the same exponential decay rate as computed in the protocol proposed earlier in Emerson \textit{et al.~}\cite{Emerson2005}, which follows from the unitary 2-design property of the Clifford group. 
This connection is made more explicit through  the observation that a random sequence of 
gates drawn from any group is equivalent to an independent sequence of twirls 
under that group, as shown explicitly in \cite{Levi2007} and had been conjectured earlier in 
\cite{Emerson2005}. 

\vspace{1mm}
\begin{minipage}[b!]{0.95 \linewidth}
	\raggedright
	\begin{protocol}[NIST RB, as described in \cite{Knill2008}]
		\hrule
		\vspace{1mm}
		~\label{proto:nist_rb}
		\vspace{0.5mm}
		\begin{enumerate}[label=\textbf{\arabic*}.]
			\hrule
			\item Sample a set of $m$ gates $G_i$ picked independently and uniformly at random from the NIST gate-set $\bf N$ defined in \cref{eq:nist_gateset};
			\item[\bf 2--8.] Idem as in \cref{proto:cliff_rb}.
			\item[\bf 9.] Fit the estimated recovery probabilities to the decay model
			\begin{align}
			A_{\bf N}~ p_{\bf N}^m + B_{\bf N}~. \label{eq:nist_decay_model}
			\end{align}
			\item[\bf 10.] Estimate the NIST gate-set infidelity through 
			\begin{align}
			r_{\bf N}= (1-p_{\bf N})/2~. \label{eq:nist_infidelity}
			\end{align}
			\hrule
		\end{enumerate}
	\end{protocol}
	\vspace{0.5mm}
\end{minipage}

Collectively these papers define what is now known as Clifford-group RB, 
 an efficient  and practical method for assessing error rates for quantum processors on arbitrarily large numbers of qubits, summarized here as Protocol \ref{proto:cliff_rb}. 
This Clifford-group RB protocol has become a \emph{de facto} standard for benchmarking and optimizing gate performance 
and has been implemented by a large number of groups across various 
hardware platforms to characterize single- and multi-qubit gate operations, see, \emph{e.g.}, Refs~\cite{xia2015,muhonen2015,Kelly2014,Barends2014,mckay2017three,sheldon2016,tarlton}.

The theoretical underpinnings of the standard protocol were clarified 
and  further developed  by Magesan \textit{et 
	al.}~\cite{Magesan2011, Magesan2012a},  which showed that the 
exponential decay rate was robust to state preparation and measurement 
errors (SPAM), and by Wallman \cite{Wallman2018GD} and Dugas \textit{et 
	al.}~\cite{carignan2018randomized}, which showed that the 
exponential decay rate was meaningfully related to a gate-fidelity in spite 
of the gauge freedom highlighted by Proctor \textit{et 
	al.}~\cite{Proctor2017} that occurs in the usual definition of  the average gate-fidelity.  
	Additionally, the work of Wallman 
\cite{Wallman2018GD}  established that the RB error rate is robust to 
very large variations in the error model over the gate set (known as 
gate-dependent error models) and thus established that RB can also be an 
effective tool for diagnosing non-Markovian errors. This follows from the fact that only 
non-Markovian errors  (including what are sometimes called 
time-dependent Markovian errors) can produce a statistically significant 
deviation from an exponential decay under a Clifford-group RB 
experiment.    
  
A different version of the 2005 Emerson \textit{et 
	al.~}\cite{Emerson2005} protocol was proposed by  Knill \textit{et 
	al.}~\cite{Knill2008} in 2008 and implemented in the NIST ion trap.  
This proposal involved the same kind of motion reversal experiment 
proposed in Emerson et al~\cite{Emerson2005} but selects a random 
sequences of gates drawn from a \textit{non-uniform} sampling of the 
single-qubit Cliffords, defined as ``Pauli-randomized $\pi/2$ gates". 
The precise recipe for this protocol is summarized as Protocol 
\ref{proto:nist_rb}.  The NIST version of the randomized benchmarking 
protocol continues to be implemented mainly in ion traps 
~\cite{brown2011single,Harty2014}. We note that in contrast to the earlier Clifford-group RB protocol which is defined for single- and multi-qubit gate operations,  the NIST version of RB is defined  only for single-qubit  gate operations. 
In this Letter, we prove that the measured fidelity under the NIST protocol will follow an exponential decay, which has never been established for this protocol, and relate the decay 
rate to the intrinsic properties of the error model, demonstrating how it differs from the properties measured by Clifford-group RB.
This analysis also provides first step towards developing a self-consistent 
theoretical framework for interpreting and relating the results of  
the large and growing family of RB-style protocols, which all share the  
structure  of applying random sequences of gates and differ mainly 
through the choice of which random gate-sets \cite{Dugas2015, Cross2016, Wallman2016a, wood2018, Chasseur2015, Wallman2015b, Gambetta2012, barends2014rolling, proctor2018direct, Kimmel2014, Emerson2007, Magesan2012a, sheldon2016, Wallman2018GD}. 

Finally, an additional motivation for the present work comes from the recent  
conceptual development \cite{wallman2016RC} establishing how accurately RB error estimation methods can inform 
the design and `in vivo' performance of large-scale quantum computations.  This development overcomes a 
standing criticism of RB  protocols that the very nature of a 
randomization protocol limits these protocols to detect only the stochastic 
component of coherent errors - and hence that RB-type protocols are not 
able to capture the full impact of these errors. Coherent errors are those that typically arise 
from  imperfect quantum control due to residual mis-calibrations 
\footnote{Note that cross-talk is a non-trivial coherent error that 
	results from control errors affecting distant qubits.} and pose a 
major challenge for reliable quantum computation. 
However, this perceived limitation has become a strength of RB 
protocols thanks to the concept of randomized compiling 
\cite{wallman2016RC}. Randomized compiling is an important 
generalization and improvement to the concept of Pauli-Frame 
Randomization (PFR) proposed earlier in \cite{knill2005nature} that does not require any overhead for the randomization and works for universal gate sets \footnote{In particular, relative to PFR, randomized compiling (i) does 
	not add additional overhead to each clock cycle, which it achieves by `compiling in' 
	the randomizing gates, (ii) works for universal gate sets, and (iii) rigorously 
	characterizes how close the effective error model is to a 
	purely stochastic error model under errors gate-dependent errors}. When implementing a quantum 
algorithm via randomized compiling, the only performance limiting 
component of a coherent error is  precisely the stochastic component 
that is detected via RB protocols.  In summary, a precise and accurate understanding of RB error estimates is highly relevant because 
RB detects precisely the component of the error that determines
 the `in vivo' performance of the gate operations within a  large-scale circuit performed via randomized compiling.

\section{Results}

\subsection{Standard RB vs NIST RB}

The standard RB protocol
(SRB)~\cite{Emerson2005,dankert2006exact} is summarized in \cref{proto:cliff_rb}. 
The recovery operations mentioned in step $2$ is usually an inversion gate, where $G_{m+1}= G_{m:1}^{-1}$, in which case 
the recovery observable simply corresponds to the initial state: $Q \approx |0 \rangle \langle 0 |$. However, performing the inverse only up to a random bit flip (i.e $G_{m+1}= X_{\pi }^b G_{m:1}^{-1}$) leads to a simpler decay model with less free parameters because then 
$B=1/2$. Of course in this case one has to keep track of the bit flip, 
that is $Q \approx X_{\pi}^b (|0 \rangle \langle 0 |)$. 
Such a randomized recovery operation
was proposed originally in \cite{Knill2008}.

SRB is typically implemented using the Clifford group ${\bf C}$ as a randomizing
gate-set, as specified in the first step of \cref{proto:cliff_rb}, 
but the derivation of the decay model shown in \cref{eq:cliff_decay_model}
holds for any unitary 2-design \cite{dankert2006exact}.
The Clifford group is defined as follows. First consider 
the pulses along any Cartesian axis system
\begin{align*}
X_\theta:=e^{-i\theta/2~\sigma_X}, ~Y_\theta:=e^{-i\theta/2~\sigma_Y}, ~Z_\theta:=e^{-i\theta/2~\sigma_Z},
\end{align*}
where $\sigma_i$ denote the unitary Pauli matrices.
The Pauli group $\bf P$ is defined in terms of the identity operation and 3 elementary $\pi$ 
pulses:
\mbox{$
\text{\bf P} :=\{\mbb I,  X_\pi, Y_\pi, Z_\pi\}.
$}
The Clifford group $\bf C$ is defined as the normalizer of the Pauli group and can be obtained from the Pauli group
composed with the coset \mbox{$\text{\bf S}:= \{\mbb I, X_{\pi/2},Y_{\pi/2},Z_{\pi/2}, Z_{\pi/2}X_{\pi/2}, X_{-\pi/2}Z_{-\pi/2}\}$}:
\begin{align}
    \text{\bf C}:= \text{\bf S} \cdot \text{\bf P}= \{ S \circ  P~ |~ S \in \text{\bf S},  P \in \text{\bf P} \}~. \label{eq:clifford_gateset}
\end{align}

Some other experimental groups performed RB using alternate $2$-design gate-sets
in step 1 of \cref{proto:cliff_rb}\cite{Barends2014}.
Amongst the set of possible unitary $2$-designs, it is worth mentioning those following subsets of {\bf C}. Consider the 
cyclic group \mbox{$\text{\bf T}:=\{\mbb I,  Z_{\pi/2}X_{\pi/2}, X_{-\pi/2}Z_{-\pi/2}\}$}, then the following sets both form
2-designs of order $12$:
\begin{align}
    {\bf C}_{12} &:= {\bf T} \cdot {\bf P}= 
    \{ T \circ  P~ |~ T \in {\bf T},  P \in {\bf P} \}~,\\
     {\bf \sqrt{Z}C}_{12} &:= Z_{\pi/2} \cdot  \text{\bf C}_{12} = 
    \{ Z_{\pi/2} \circ  C~ |~ C \in {\bf C}_{12} \}~,
\end{align}
with ${\bf C}_{12} \cup \bf{\sqrt{Z}C}_{12} = \bf C$.
Obviously, the decay parameters as well as the infidelity
depend on the randomizing gate-set
(hence the indices).
The validity of the decay model and the connection between the decay parameter and the gate-set infidelity have been 
demonstrated in the case of gate-independent Markovian noise scenarios
in \cite{Emerson2005}. The proofs of \cref{eq:cliff_decay_model} and \cref{eq:cliff_infidelity}
have been generalized to encompass gate-dependent noise scenarios 
in \cite{Wallman2018GD,Merkel2018} and \cite{carignan2018randomized} respectively\footnote{In gate-dependent noise scenarios, the connection between the RB decay parameter and 
the gate-set infidelity remains a (strongly supported) conjecture for $d >2$.}.
Although the proof techniques can get mathematically heavy, their essence remains simple:
the algebraic richness of $2$-designs
prevents errors to accumulate in an unpredictable way
as the circuit grows in length. 
As we show with more care in the next section, 
the random sampling over 
the gate-set tailors the effective errors 
at each cycle in a depolarizing channel
for which the evolution is parameterized by a single
real number $p$. The errors are stripped out of 
all their properties except one, which turns out
to be in one-to-one correspondence with their average infidelity.
By modifying the sampled circuits lengths, we can estimate 
the parameter $p$ and retrieve the infidelity.

While unitary $2$-designs are provably effective randomizing gate-sets, leading to the model portrayed in \cref{eq:cliff_decay_model}, 
some algebraically weaker gate-sets have indicated a similar exponential decaying behaviour.

The gate-set {\bf N} used in NIST RB \cite{Knill2008} is a composition of a set $
\text{\bf Q} :=\{ X_{\pm\pi/2},  Y_{\pm \pi/2}\},
$ 
consisting of $\pi/2$ pulses in the xy-plane,
with the Pauli operators:
\begin{align}
    {\bf N}:= {\bf Q} \cdot {\bf P} =      \{  Q \circ  P~ |~ Q \in {\bf Q},  P \in {\bf P} \}~. \label{eq:nist_gateset}
\end{align}
{\bf N} has order 8, and although it contains all its inverse elements 
(that is $\forall  N \in \bf N$, $\exists  M \in \bf N$  s.t. 
$ M \cdot  N = \mbb I $), it is not closed under multiplication. It 
does not form a group, nor a 2-design; however, the closure $\langle 
\bf N \rangle$ forms the Clifford group $\bf C$. 

\begin{figure}[h!]
	
	\includegraphics[width=1.0\linewidth]{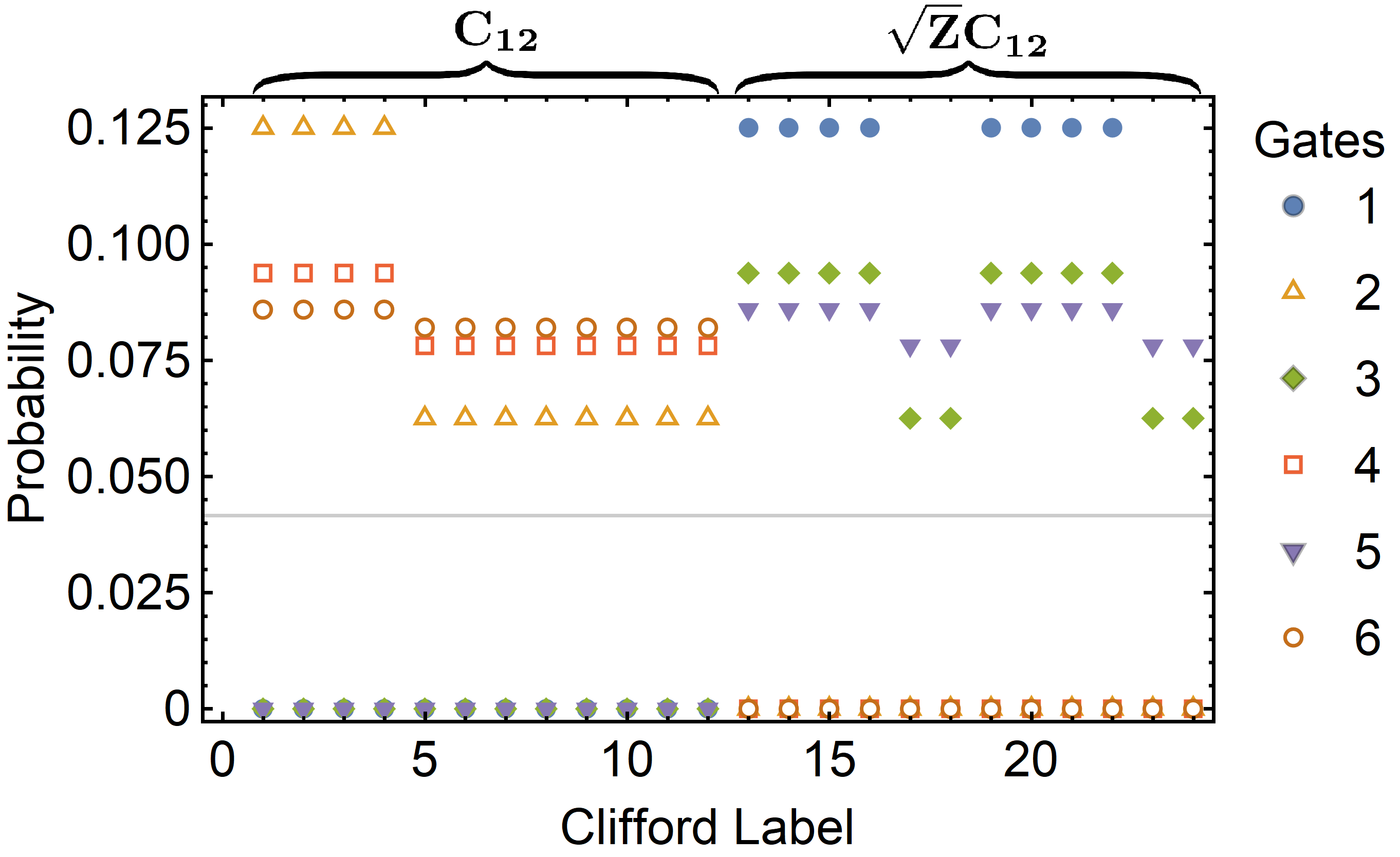}
	\caption{Probability distribution over the Clifford gates {\bf C}
		(labelled as in \cite{Barends2014}) after $m$ 
		gates (i.e clock cycles) of NIST RB drawn uniformly at random from 
		${\bf N} \subset {\bf C}$. This leads to a \emph{non-uniform} sampling over the 
		Cliffords that varies as $m$ increases. 
		Asymptotically, for a sequence of even (or odd) length, the probability distribution tends toward a uniform distribution over 
		$\bf{C}_{12}$ (or $\bf{\sqrt{Z}C}_{12}$).
		The grey line indicates an equal probability over the full 24 Clifford group $\bf C$.}
	\label{fig:hist}
\end{figure}

RB sequences can be seen as Markov chains \cite{meier}, where the elements of the chain are the aggregate circuits, that is $C_1= G_1$, $C_2= G_2 G_1$, $C_m= G_{m:1}$. Indeed, the probability distribution on circuits $C_m=G_{m:1}$ simply depends on the circuit $C_{m-1}$ and on the probability distribution of the random gate applied at step $m$.  In standard RB, $C_i$ is always uniformly distributed over the Clifford group. In NIST RB, $C_{2n}$ (or $C_{2n+1}$) converges to a uniform distribution over $\bf{C}_{12}$ (or $\bf{\sqrt{Z}C}_{12}$), as shown in \cref{fig:hist}.

While this approach to RB has been useful for estimating 
error rates \cite{Knill2008,brown2011single,Harty2014}, in the 
absence 
of a unitary 2-design structure, it is not clear how
to relate the measured probabilities from 
\cref{proto:nist_rb}
to the usual decay predicted under SRB, 
or to any infidelity for that matter. 
In this paper we provide a concrete analysis
of the outcome of \cref{proto:nist_rb}, 
which yields a justification and interpretation for the decay model
\cref{eq:nist_decay_model,eq:nist_infidelity}.
 
It is important to emphasize that NIST RB 
now falls into a family of RB protocols defined as  ``direct RB'' \cite{proctor2018direct}. The analysis below gives a concrete instance 
of direct RB that both justifies and interprets past experiments and gives an insightful example
of the main idea behind direct RB. 

\subsection{Theoretical Analysis of NIST RB}\label{analysis}

The goal of this section is to provide the key insight
behind the mechanics of NIST RB. To lighten up the mathematical machinery, 
we assume a gate-independent error model\footnote{The formal analysis of direct RB under more general gate-dependent noise scenarios will be considered in subsequent work. However, in \cref{sec:gate_dependent} we elucidate the broad reasoning
required for a gate-dependent analysis.}, where the noisy gates are 
followed by an error $\Lambda$:
\begin{align}
\tilde{G} = \Lambda G~.
\end{align}
In such model, the gate-set infidelities $r_{\bf C}$ and $r_{\bf N}$
are \emph{de facto} equal to the infidelity of the error $r(\Lambda, \mbb I)$. 
We proceed in showing that $r(\Lambda,\mbb I)$ can be estimated by both 
\cref{proto:cliff_rb,proto:nist_rb}.

The recovery probabilities look like 
\begin{align}
    \tr Q \Lambda \underbrace{X_{\pi}^b G_{m:1}^{-1}}_{G_{m+1}}\Lambda G_m \cdots \Lambda G_2 \Lambda G_1(\rho)~.
\end{align}
Shoving the last error $\Lambda$ as well as the random bit flip $X_{\pi}^b$ in the measurement 
procedure (that is, $Q 
\rightarrow  X_{\pi}^b\Lambda^{\dagger}(Q)$), leaves us 
with the random sequence which is at the heart of both NIST RB and SRB protocols:
\begin{align}
    S(\{G_i\}) = G_{m:1}^{-1}\Lambda G_m \cdots \Lambda G_2 \Lambda G_1~.
\end{align}

In SRB, the next step in the 
analysis consists in redefining the gates as $G_i =  
G'_i G_{i-1}'^{-1}$ (with $G_1=G_1'$), where both $G_i$ 
and $G_i'$ are picked uniformly at random from the randomizing set. 
Such a relabeling is 
possible because the randomizing gate-set is usually a group. Averaging over
all sequences yields
\begin{align}
     \mbb E_{\{G_i\}} S(\{G_i\}) &= \mbb E_{\{G'_i\}} G_{m}'^{-1}\Lambda G_m' \cdots G_2'^{-1} \Lambda G_2' G_1'^{-1} \Lambda G'_1~\notag \\
     &= \left(\Lambda^{\bf C}\right)^m~,
\end{align}
where
\begin{align}
    \Lambda^{\bf C}:= \frac{1}{|\bf C|} \sum_{G \in {\bf C}} G^{-1} \Lambda G \label{eq:twirl}
\end{align}
is referred to as the twirl of the error $\Lambda$ over
the gate-set $\bf C$. If $\bf C$ is a $2$-design, then the twirled channel
$\Lambda^{\bf C}$ is reduced to a depolarizing channel. 
To mathematically concretize the description of a channel $\Lambda$, 
we resort to the $4 \times 4$ Pauli-Liouville 
representation, which is defined as
\begin{align}
    \Lambda_{ij}:=\frac{1}{2} \tr B_j^\dagger \Lambda(B_i)
\end{align}
where $B_1= \mbb I$, $B_2= \hat{\sigma}_x$, $B_3= \hat{\sigma}_y$, $B_4= \hat{\sigma}_z$.
In such representation, the depolarizing channel $\Lambda^{\bf C}$
is expressed as a diagonal matrix
\mbox{${\rm diag}(1 ,  p_{\bf C} ,  p_{\bf C} ,  p_{\bf C})$}, where $ p_{\bf C}$ is a real number close to 1:
\begin{align}
     p_{\bf C}= \frac{\Lambda_{22}+\Lambda_{33}+\Lambda_{44}}{3}~. \label{eq:p_c_liouville}
\end{align}

The averaged core sequence hence evolves as
\begin{align}
    \mbb E_{\{G_i\}} S(\{G_i\}) = \left(\Lambda^{\bf C}\right)^m={\rm diag}(1 ,   p_{\bf C}^m ,  p_{\bf C}^m ,  p_{\bf C}^m)~. \label{eq:avg_seq_srb}
\end{align}
Deriving \cref{eq:cliff_decay_model} is then simply a matter of incorporating 
SPAM procedures in the evaluation of the recovery probabilities. 
Straightforward algebra links the infidelity of $\Lambda$
with its diagonal Liouville matrix elements through
\begin{align}
    r(\Lambda,\mbb I)=  \frac{1}{2}- \frac{\Lambda_{22}+\Lambda_{33}+\Lambda_{44}}{6}~. \label{eq:inf_liouville}
\end{align}
The relation between the decay constant $ p_{\bf C}$ and the gate-set infidelity $r_{\bf C}=r(\Lambda,\mbb I)$ results from combining \cref{eq:p_c_liouville} and \cref{eq:inf_liouville}.

The relabeling trick resulting in a $m$-composite depolarizing channel
is not possible in NIST RB: $\bf N$ is neither a group nor a $2$-design. 
However, although $\bf N$ has a weaker algebraic structure, 
it is not completely devoid of interesting properties. Indeed, every 
element of $\bf N$ can be written as $ P_{\rm left} \cdot  Q 
\cdot  P_{\rm right}$, where $ P_{\rm left},~ P_{\rm right} \in 
{\bf P}$ and $ Q \in {\bf Q}$. Using this, we can relabel every gate 
$G_i$ as 
\begin{subequations}
	\begin{align}
	G_1 & =  P_1  Q_1~, \\
	G_i &=  P_i  Q_i  P_{i-1}^{-1} ~~(i=2, \cdots, m)~,\\
	G_{m:1}^{-1} & =   Q_{m:1}^{-1}  P_m^{-1}
	\end{align}
\end{subequations}
where $ P_i$ is chosen UAR from the Pauli group $\bf P$, and $ 
Q_i$ are 
chosen UAR from $\bf Q$. Using such a manipulation and randomizing over 
the Paulis transform the core sequence into
\begin{align} \label{eq:PQ}
\mbb E_{\{ P_i\} } S(\{G_i\})= 
Q_{m:1}^{-1}\Lambda^{\bf P}  Q_m \cdots \Lambda^{\bf P}  Q_2 
\Lambda^{\bf P}  Q_1~,
\end{align}
where $\Lambda^{\bf P}$ is the error channel twirled over the Pauli 
group. In the Pauli-Liouville picture, the Pauli group has $4$ 
inequivalent irreps; the twirled channel is diagonal:
\begin{align} 
\Lambda^{\bf P}= {\rm diag}(1 , x , y , z) ~, 
\end{align}
 where $x=\Lambda_{22}$, $y=\Lambda_{33}$, 
$z=\Lambda_{44}$. The relabeling method still can't be used with the 
$ Q_i$'s, but the simplification of the noise channel $\Lambda$ 
through the Pauli twirl unveils a recursive  approach. Consider the 
$m=1$ case: 
\begin{align} 
\mbb E_{\{G_1\}} S(\{G_{1}\})=\mbb E_{\{ Q_1\} }  Q_{1}^{-1}\Lambda^{\bf P}  Q_1= \Lambda^{\bf N} 
~ ,
\end{align}
where the twirl over the NIST gate-set results in 
\begin{align} 
\Lambda^{\bf N}= {\rm diag}\left(1, \frac{x+z}{2} ,\frac{y+z}{2} , 
\frac{x+y}{2}\right) ~. \label{eq:nist_twirl}
\end{align}
The $m=2$ case suggests a recursion relation:
\begin{subequations}
	\begin{align} 
	\mbb E_{\{G_{i}\} } S(\{G_{i}\}) &= 
	\left(\Lambda^{\bf N} \Lambda^{\bf P}\right)^{\bf N} ~ , \\
	\left(\Lambda^{\bf N} \Lambda^{\bf P}\right)^{\bf N} &= {\rm diag} (1, 
	x_2 ,y_2 , z_2)~,
	\end{align}
\end{subequations}
where
\begin{subequations}
	\begin{align} 
	x_2 = \frac{x\frac{(x+z)}{2}+z\frac{(x+y)}{2}}{2}~ , \\
	y_2 = \frac{y\frac{(y+z)}{2}+z\frac{(x+y)}{2}}{2}~ , \\
	z_2 = \frac{x\frac{(x+z)}{2}+y\frac{(y+z)}{2}}{2}~ . 
	\end{align}
\end{subequations}
Indeed, the general case can be expressed as
\begin{subequations}
	\begin{align} 
	&\mbb E_{\{G_{i}\} } S(\{G_{i}\}) = 
	\left(\left(\left(\Lambda^{\bf N} \Lambda^{\bf P}\right)^{\bf N} 
	\Lambda^{\bf P}  \right)^{\bf N} \Lambda^{\bf P} \cdots 
	\right)^{\bf N}~ , \\
	& \left(\left(\left(\Lambda^{\bf N} \Lambda^{\bf P}\right)^{\bf 
		N} \Lambda^{\bf P}  \right)^{\bf N} \Lambda^{\bf P} \cdots 
	\right)^{\bf N}= {\rm diag}(1, x_m , y_m , z_m)~,
	\end{align}
\end{subequations}
where the recursion relation can be stated as
\begin{subequations}
	\begin{align} 
	x_m = \frac{x \cdot x_{m-1}+z \cdot z_{m-1}}{2}~ , \\
	y_m = \frac{y\cdot y_{m-1}+z\cdot z_{m-1}}{2}~ , \\
	z_m = \frac{x \cdot x_{m-1}+y \cdot y_{m-1}}{2}~ . 
	\end{align}
\end{subequations}
Using basic linear algebra, this system of recursive equations can be expressed as
\begin{align}
\begin{bmatrix}
x_m \\
y_m \\
z_m
\end{bmatrix}
&=
M \begin{bmatrix}
x_{m-1} \\
y_{m-1}  \\
z_{m-1} 
\end{bmatrix}= M^m \begin{bmatrix}
1 \\
1 \\
1 
\end{bmatrix}~, \label{eq:mat_recursion}
\end{align}
where 
\begin{align}
M=\frac{1}{2} 
\begin{bmatrix}
x& 0& z \\
0 & y & z \\
x & y& 0
\end{bmatrix}~. \label{eq:M}
\end{align}
$x$,$y$ and $z$ differ from $1$ by at most order $r(\Lambda,\mbb I)$. 
Hence, up to the second order in the infidelity, $M$ has the following spectrum:
\begin{subequations}
	\begin{align} 
	\lambda_1 & \approx \frac{x+y+z}{3} = p_{\bf C}~, \label{eq:lambda_1} \\
	\lambda_2 & \approx \frac{x+y}{4}~, \\
	\lambda_3 & \approx -\frac{x+y+4z}{12} ~.
	\end{align}
\end{subequations}
Since $\lambda_1 \approx 1$, $\lambda_2 \approx 1/2$ and $ \lambda_3 \approx -1/2$, 
$M^m$ converges very quickly to a rank-$1$ operator as $m$ increases. This means that for $m$ large enough so that $1/2^m$ becomes negligible, $x_m$, $y_m$, $z_m$ are proportional to $\lambda_1^m$:
\begin{align} 
	\mbb E_{\{G_{i}\} } S(\{G_{i}\}) \approx {\rm diag}(1, c_1 \lambda_1^m , 
 c_2 \lambda_1^m,c_3 \lambda_1^m)~ , \label{eq:avg_seq_nist}
\end{align}
where $c_i$ are proportionality constants. \Cref{eq:nist_decay_model}
is obtained by incorporating the SPAM procedures in evaluating the 
recovery probabilities, and by relabeling $\lambda_1$ as $p_{\bf N}$. 
Finally, the relation between the decay $p_{\bf N}$ and the 
gate-set infidelity $r_{\bf N}= r(\Lambda,\mbb I)$ is retrieved via
\cref{eq:lambda_1}:
\begin{align}
    r_{\bf N}= (1-p_{\bf N})/2 +O(r^2_N)~,
\end{align}
which essentially states that the NIST RB decay parameter $p_{\bf N}$
provides a very good estimates of the gate-set infidelity $r_{\bf N}$ through
\cref{eq:nist_infidelity}.

With this analysis behind us, let's compare the internal mechanics
of \cref{proto:cliff_rb,proto:nist_rb}. First of all, 
both protocols make use of randomizing gate-sets, $\bf C$ and 
$\bf N$ respectively. In both cases, the randomization
tailors the error dynamics such that
the average core sequence $\mbb E_{\{G_i\}} S(\{G_i\})$ 
evolves with respect to a single decay parameter, as show in 
\cref{eq:avg_seq_srb,eq:avg_seq_nist}. An interesting difference 
here is that the Clifford randomization 
simplifies the error into a $1$-parameter depolarizing channel
at each time step, while the NIST randomization doesn't, 
as shown in \cref{eq:nist_twirl}. In the latter case, 
certain error components
remain ``imperfectly shuffled'' after a few 
random gates, leaving space for a multi-parameterized noise
evolution portrayed by \cref{eq:mat_recursion,eq:M}. However,
as the random sequence gets longer, 
the evolution quickly converges to a $1$-parameter decay.
The fact that this decay relates to the infidelity shouldn't be
surprising, since ${\rm diag}(0,1,1,1)$ is a channel component
that commutes with every unitary (so is ``immuned'' to twirling). Given an 
error $\Lambda$, its corresponding coefficient is $(\Lambda_{22}+\Lambda_{33}+\Lambda_{44})/{3}$,
which is in one-to-one correspondence with the infidelity $r(\Lambda,\mbb I)$
via \cref{eq:inf_liouville}.

\subsection{Measured Error Rates under NIST RB vs SRB}

\begin{figure*}
	\centering
	\begin{minipage}{0.66\columnwidth}
		\centering
		\includegraphics[width=1.0\columnwidth]{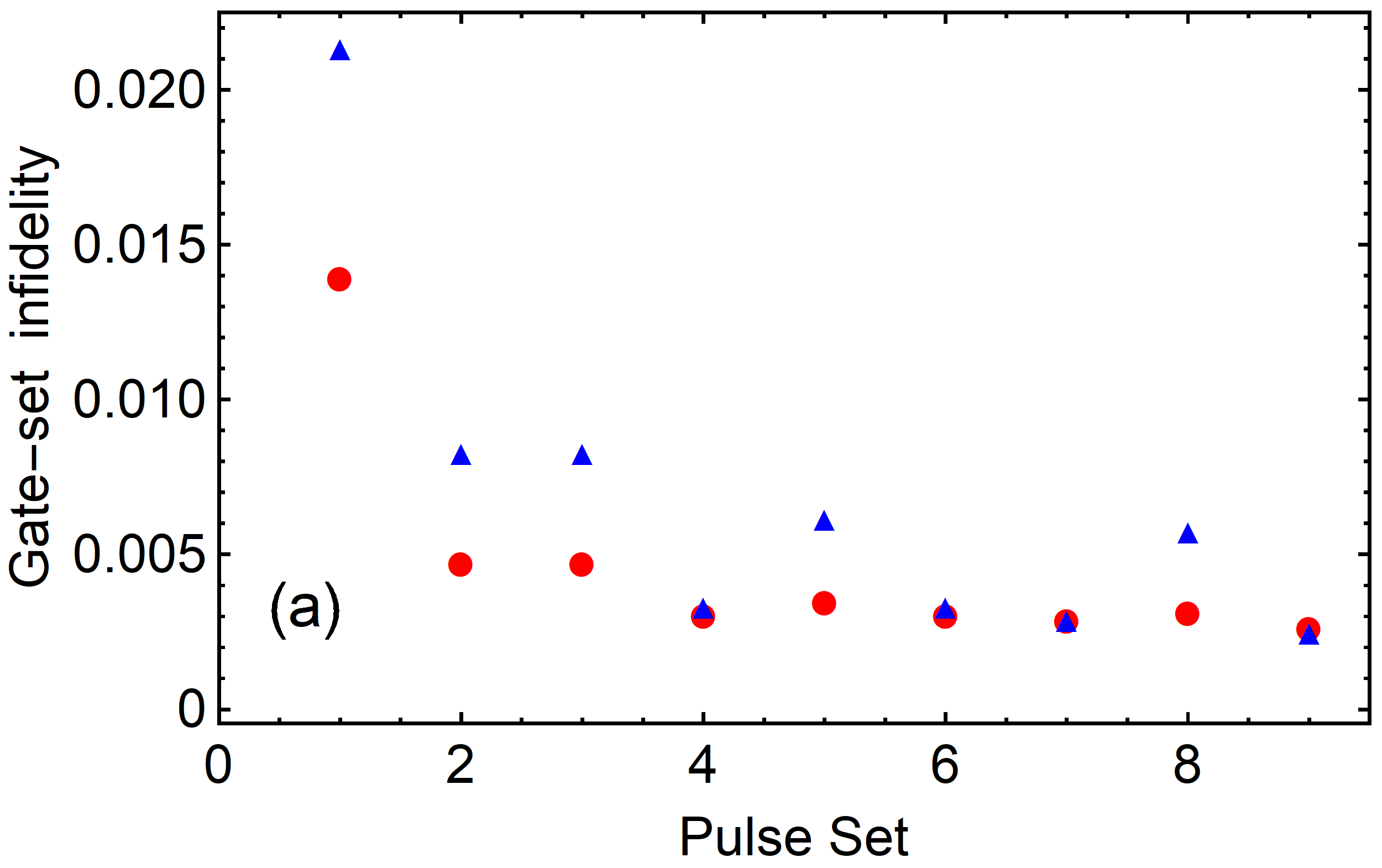}
		\label{fig:CvsNO}
	\end{minipage}\hfill
	~
	\begin{minipage}{0.66\columnwidth}
		\centering
		\includegraphics[width=1.0\columnwidth]{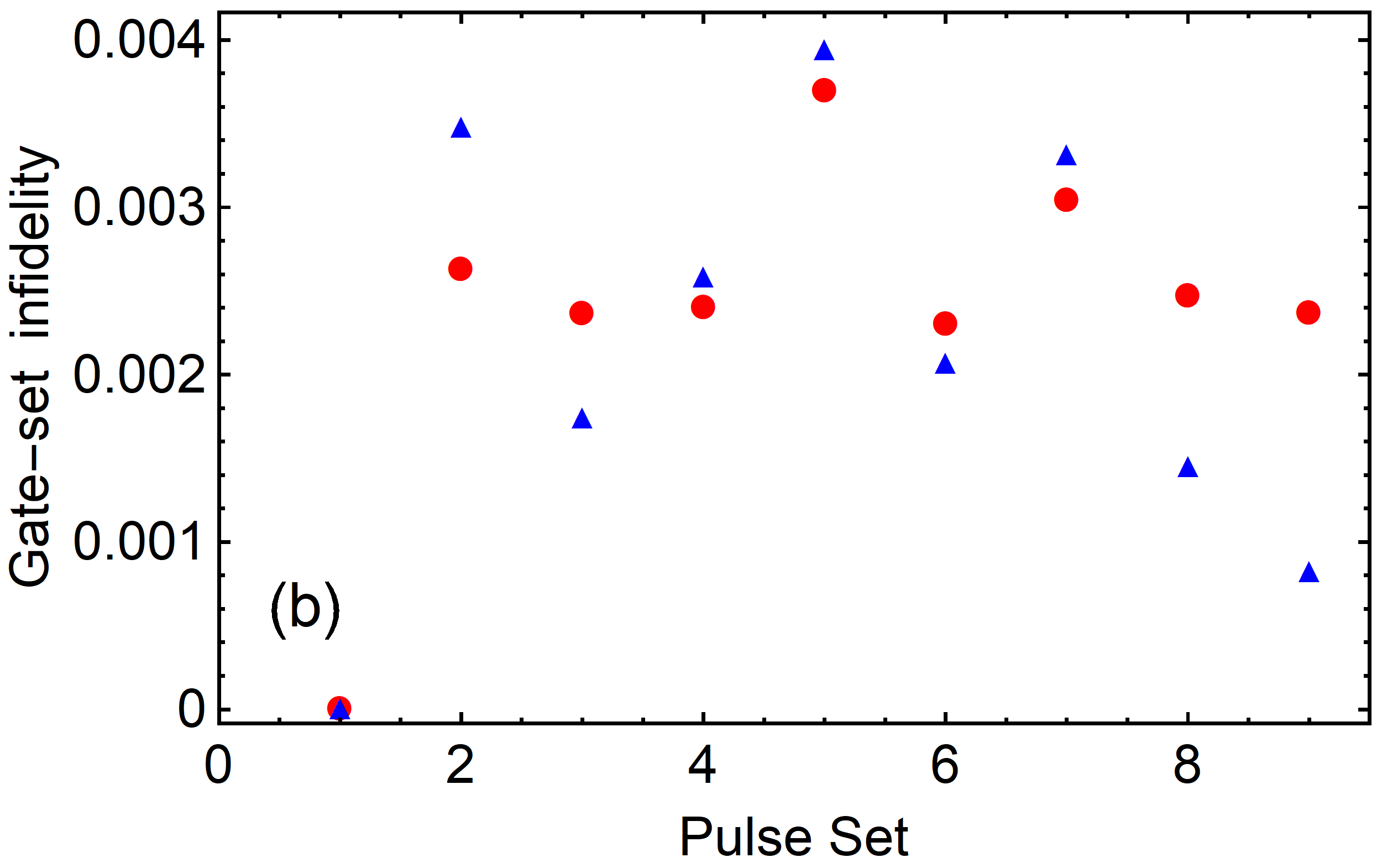}
		\label{fig:CvsNZ}
	\end{minipage}\hfill
	~
	\begin{minipage}{0.66\columnwidth}
		\centering
		\includegraphics[width=1.0\columnwidth]{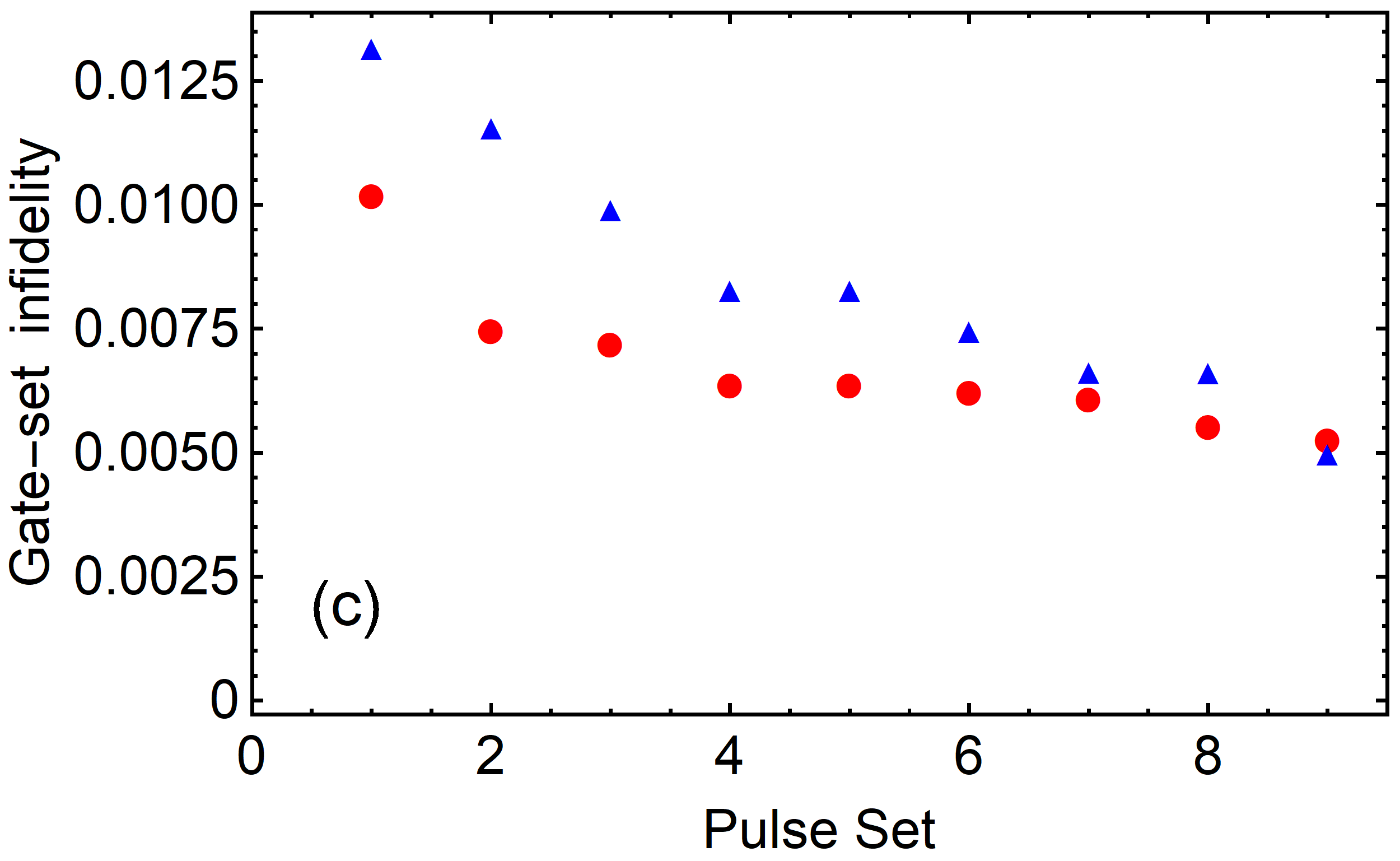}
		\label{fig:CvsNP}
	\end{minipage}
	\caption{(Color online) A comparison between the gate-set infidelities $r_{\bf N}$  (blue triangles) and $r_{\bf C}$ (red points) for pulse sets from \cref{table:gg} with gate-dependent (a) coherent over-rotation error $\tilde{\mbb I}=\mbb I$, 
		$\tilde{X}_{\pm\theta}=X_{\pm(\theta+0.1)}$, 
		$\tilde{Y}_{\pm\theta}=Y_{\pm(\theta+0.1)}$, 
		$\tilde{Z}_{\pm\theta}=Z_{\pm(\theta+0.1)}$ 
		(b) coherent 
		Z-rotation error $\tilde{\mbb I}=Z_{0.1}$, 
		$\tilde{X}_{\theta}=Z_{0.1} X_{\theta}$, 
		$\tilde{Y}_{\theta}=Z_{0.1} Y_{\theta}$, 
		$\tilde{Z}_{\theta}=Z_{\theta + 0.1}$, 
		and (c) incoherent dephasing error  $\tilde{\mbb I}= D_{0.99}$, $\tilde{X}_\theta= D_{0.99}X_\theta,~\tilde{Y}_\theta= D_{0.99}Y_\theta,~ \tilde{Z}_\theta= D_{0.99}Z_\theta$, where
		$ D_{\alpha}={\rm diag}(1 ,  \alpha , \alpha , 1).~$ 
		Under these error models and pulse sets, $r_{\bf N}$ and $r_{\bf C}$ differ by up to a factor of 3, which could significantly affect the expected overhead under quantum error correction.}
	\label{fig:CvsN}
\end{figure*}

\begin{figure*}
	\centering
	\begin{minipage}{0.66\columnwidth}
		\centering
		\includegraphics[width=1.0\columnwidth]{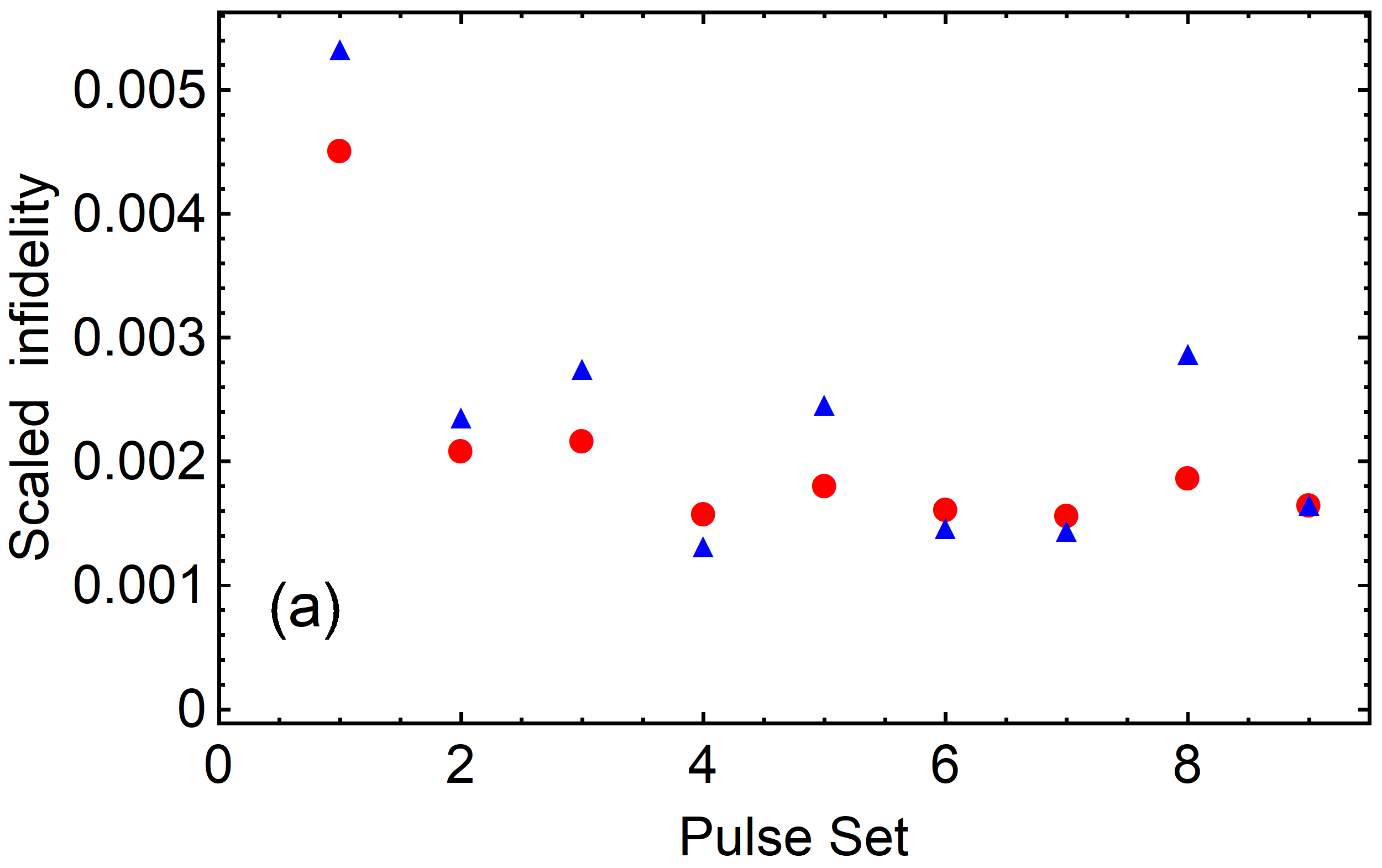}
		\label{fig:EPPO}
	\end{minipage}\hfill
	~
	\begin{minipage}{0.66\columnwidth}
		\centering
		\includegraphics[width=1.0\columnwidth]{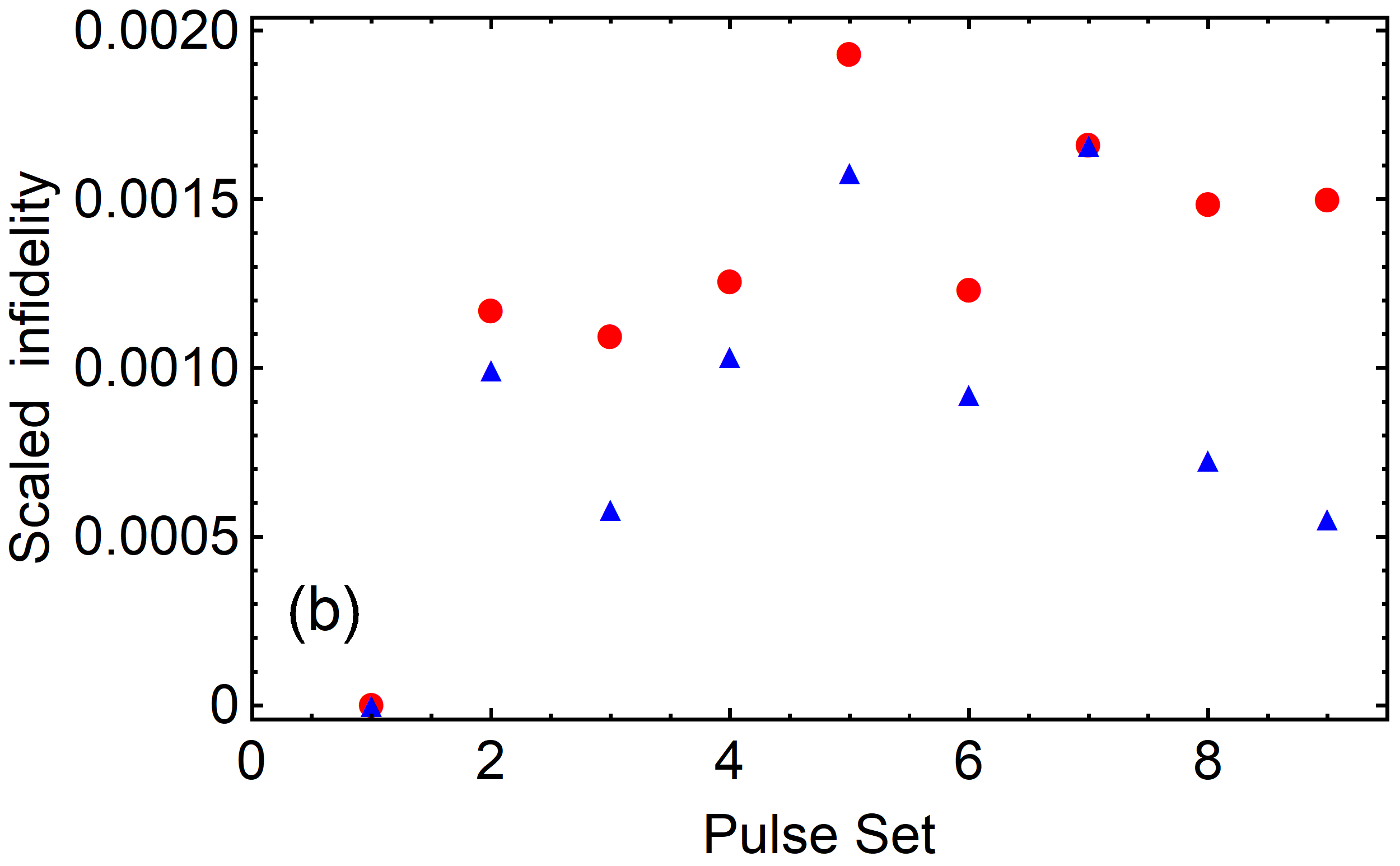}
		\label{fig:EPPZ}
	\end{minipage}\hfill
	~
	\begin{minipage}{0.66\columnwidth}
		\centering
		\includegraphics[width=1.0\columnwidth]{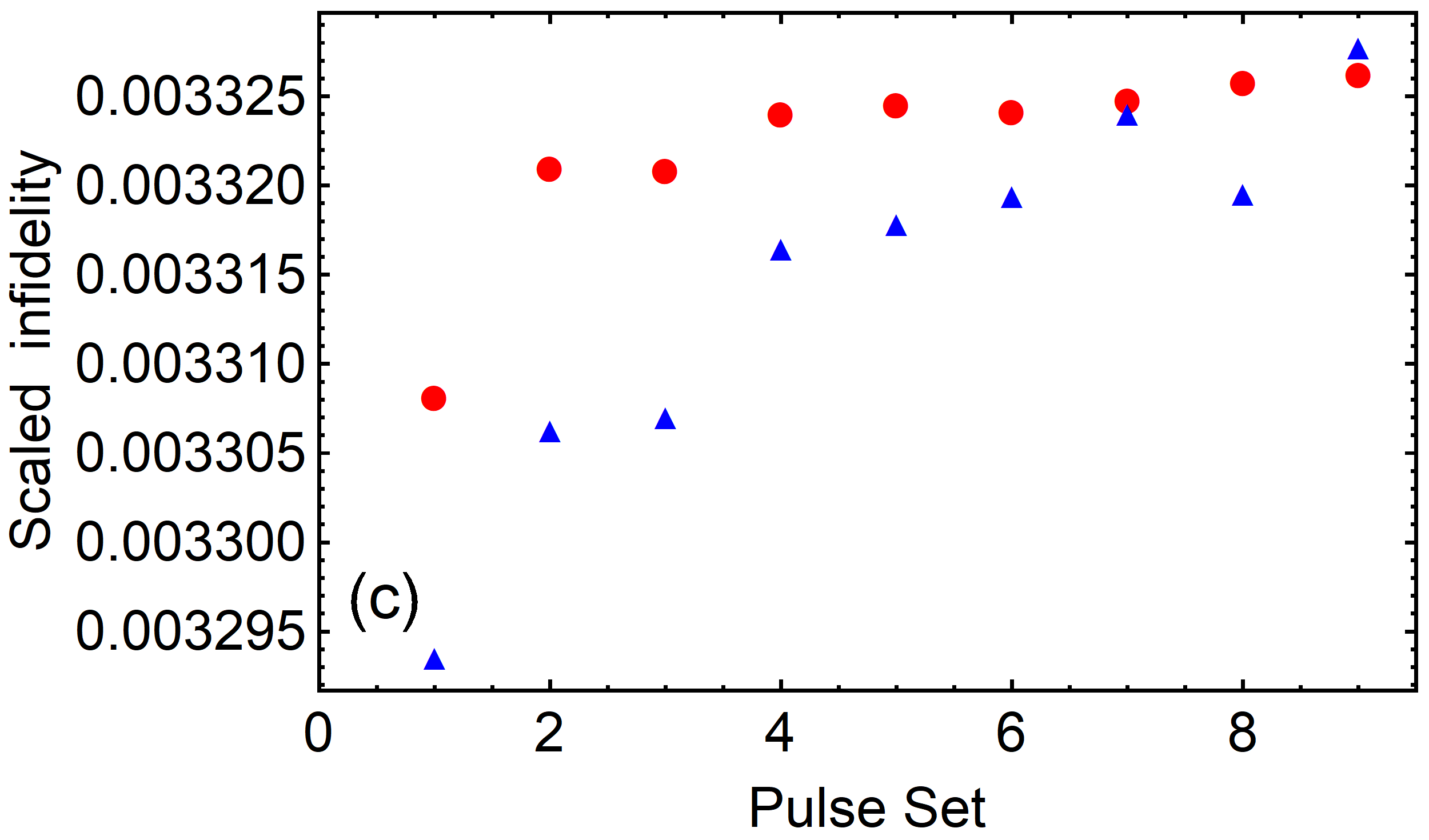}
		\label{fig:EPPP}
	\end{minipage}
	\caption{(Color online) A comparison between the scaled infidelities $r_{\bf N}/n_{{\bf N}}$ (blue triangles) and $r_{\bf C}/n_{{\bf C}}$ (red points) using pulse sets from 
		\cref{table:gg} with gate-dependent (a) coherent over-rotation error $\tilde{\mbb I}=\mbb I$, 
		$\tilde{X}_{\pm\theta}=X_{\pm(\theta+0.1)}$, 
		$\tilde{Y}_{\pm\theta}=Y_{\pm(\theta+0.1)}$, 
		$\tilde{Z}_{\pm\theta}=Z_{\pm(\theta+0.1)}$ 
		(b) coherent 
		Z-rotation error $\tilde{\mbb I}=Z_{0.1}$, 
		$\tilde{X}_{\theta}=Z_{0.1} X_{\theta}$, 
		$\tilde{Y}_{\theta}=Z_{0.1} Y_{\theta}$, 
		$\tilde{Z}_{\theta}=Z_{\theta + 0.1}$, 
		and (c) incoherent dephasing error  $\tilde{\mbb I}= D_{0.99}$, $\tilde{X}_\theta= D_{0.99}X_\theta,~\tilde{Y}_\theta= D_{0.99}Y_\theta,~ \tilde{Z}_\theta= D_{0.99}Z_\theta$, where $ D_{\alpha}= {\rm diag}(1 ,  \alpha , \alpha , 1).~$
		Clearly, even after accounting 
		for the discrepancy between the average number of pulses per 
		gate of these two RB protocols ($n_{\bf N} ~\&~ n_{\bf C}$), the 
		measured error rates still differ due to 
		their differing sampling  over the gate sets.}
	\label{fig:EPP}
\end{figure*}

\renewcommand{\arraystretch}{2}

\begin{table}[h]
	\centering
	\begin{tabular}{|P{0.8cm}|P{4.2cm}|P{1.1cm}|P{1.1cm}|}
		\hline
		
		Index	&	Pulse Set	&	$n_{\bf C}$	&	$n_{\bf N}$\\\hline
		1	&  $\{\mbb I, \tilde{X}_{+\pi/2},\tilde{Y}_{+\pi/2}\}$	&	3.08333  &	4.0\\\hline
		2	&  $\{ \tilde{X}_{\pm\pi/2},\tilde{Y}_{\pm\pi/2}\}$	&	2.25	&	3.5 \\\hline
		3	&  $\{\mbb I, \tilde{X}_{\pm\pi/2},\tilde{Y}_{\pm\pi/2}\}$	&	2.16667	&	3.0  \\\hline
		4	&  $\{\tilde{X}_\pi,\tilde{Y}_\pi, \tilde{X}_{\pm\pi/2},\tilde{Y}_{\pm\pi/2}\}$	&	1.91667	&	2.5  \\\hline
		5	&  $\{\tilde{\mbb I},\tilde{Z}_\pi, \tilde{X}_{\pm\pi/2},\tilde{Y}_{\pm\pi/2}\}$ &	1.91667	&	2.5 \\\hline
		6	&  $\{\tilde{\mbb I},\tilde{X}_\pi,\tilde{Y}_\pi, \tilde{X}_{\pm\pi/2},\tilde{Y}_{\pm\pi/2}\}$ &	1.875	&	2.25 \\\hline
		7	&  $\{\tilde{\mbb I},\tilde{X}_\pi,\tilde{Y}_\pi, \tilde{Z}_\pi, \tilde{X}_{\pm\pi/2},\tilde{Y}_{\pm\pi/2}\}$ &	1.8333	&	2.0 \\\hline
		8	&  $\{\mbb I,Z_\pi, \tilde{X}_{\pm\pi/2},\tilde{Y}_{\pm\pi/2}\}$ &	1.66667	&	2.0 \\\hline
		9	&  $\{\mbb I,\tilde{X}_\pi,\tilde{Y}_\pi,Z_\pi, \tilde{X}_{\pm\pi/2},\tilde{Y}_{\pm\pi/2}\}$	&	1.58333	&	1.5 \\\hline
		
	\end{tabular} 
	\caption{Each of the 24 $\bf C$s and 8 $\bf N$s were constructed 
		by a sequence of noisy ($\tilde{\mbb I},\tilde{X}_\theta,\tilde{Y}_\theta,\tilde{Z}_\theta$) and virtually (ideal) implemented ($\mbb I,X_\theta ,Y_\theta, Z_\theta$) pulses.
		Note that when implementing the $\pi$ pulses, the direction of rotation (sign of $\pi$) is selected uniformly at random as described in \cite{Knill2008}. $n_{\bf C}$ and $n_{\bf N}$ are the average 
		number of noisy pulses per gate from $\bf C$ and $\bf N$, and is used for calculating 
		the scaled infidelity ($r_{\bf C}/n_{\bf C}$ and $r_{\bf N}/n_{\bf N}$).}
	\label{table:gg}
\end{table}

In the previous section, we showed that under the assumption 
of gate-independent errors, the gate-set infidelities 
$r_{\bf C}$ and $r_{\bf N}$ could be estimated 
via SRB and NIST RB respectively, and that these estimates
did both coincide with $r(\Lambda, \mbb I)$. In reality, one might find
through experiment that $r_{\bf C}$ and $r_{\bf N}$ differ quite substantially. This, of course, is explained by gate-dependent effects: certain gates have
higher infidelities than others, and since $r_{\bf C}$ and $r_{\bf N}$ consist
in the expected gate infidelity \emph{over their respective gate-set} (see \cref{sec:gate_dependent} for further justifications), they will yield different values.

To demonstrate this, we numerically simulated both SRB and NIST RB experiments implemented in different fashions, using a
plethora of primitive pulse sets (see table 
\ref{table:gg}) each undergoing various physically realistic gate-dependent noise scenarios.
The respective infidelities $r_{\bf C}$ and $r_{\bf N}$ are juxtaposed in 
\cref{fig:CvsN}, and
differ by up to a factor of $\sim 3$.
This might not strike as a major difference
in a day and age where the infidelity is typically filtered 
through its order of magnitude. That being said,
for both surface and concatenated quantum error correcting codes, 
the overhead becomes more sensitive to the error rate as it approaches 
the fault-tolerant threshold.  Therefore, a factor of 3 could 
dramatically increase the overhead (by more than an order of magnitude) 
if the error rate is close to the threshold.  
In the extreme case, the factor of 3 could 
cause the error rate to surpass the threshold and 
fault tolerance would become impossible. 

One might be tempted to reconcile those 
different infidelities by accounting for the number of primitive pulses
that form each gate-set element. That is, if the Clifford gates $C \in {\bf C}$
necessitate the average application of $n_{\bf C} = 1.875$ pulses and that 
the NIST gates $N \in {\bf N}$ require an average of $n_{\bf N}= 2.25$ pulses, 
it might seem natural to scale the infidelities as $r_{\bf C}/n_{\bf C}$ and 
$r_{\bf N}/n_{\bf N}$.

This is a common misconception because scaling the infidelity is only meaningful if it grows linearly with the number of pulses, as is the case of purely stochastic error. For example, in gate-dependent dephasing scenarios, the scaled infidelities ($r_{\bf C}/n_{\bf C}$ and 
$r_{\bf N}/n_{\bf N}$) only differ by $O(r_{\bf N}^2)$ because the error is purely stochastic (incoherent) and pulse-independent (see \cref{fig:EPP}c).

Contrarily, it has been shown that under coherent error scenarios, the composite errors can vary \cite{Dugas2016}, as they can positively and negatively interfere. Therefore, when the error is coherent, NIST RB and SRB obtain different scaled infidelities, despite the infidelity per pulse remaining fixed (see figs. \ref{fig:EPP}a and \ref{fig:EPP}b) because they are sampling differently from the pulse sets, which causes them to probe different coherent error models. As such, it is bad practice to measure the scaled infidelity which is not equivalent to, and should not be confused with, the error per pulse.

\vspace{-2em}
\section{Conclusion}

Randomized benchmarking (RB) is an important tool for estimating error rates 
associated with sets of elementary gate operations. SRB and NIST RB are 
two distinct RB protocols that have been 
confused in the literature and can lead to distinct outcomes. In this work we developed a rigorous theoretical framework proving that NIST RB, like SRB,
leads to an exponential decay which depends only on the underlying 
gate-independent error model. We showed 
SRB, which samples from a uniform 2-design, and NIST RB, which samples from a subset of 
it, lead to significantly different observed error rates for a variety of 
physically realistic gate-dependent error models and pulse sets (see fig. 
\ref{fig:CvsN}). In typical cases the error rates  differ by up to a factor 
of 3, which could have a significant impact on the overhead when 
implementing fault-tolerant quantum error correction. 

A next step is to develop a rigorous theoretical 
framework under which the experimental results from NIST RB and other RB methods using arbitrary gate 
sets can be analyzed in a unified and related to infer properties of the underlying error model in a  consistent manner.
 
\vspace{-1em}
\subsection*{Acknowledgements}
We thank Daniel Gottesman for fruitful discussions on QECC. This research was
supported by the U.S. Army Research Office through
grant W911NF-14-1-0103. This research was undertaken
thanks in part to funding from TQT, CIFAR, the Government
of Ontario, and the Government of Canada through
CFREF, NSERC and Industry Canada.

\vspace{-1em}
\appendix
\section{NIST RB analysis under gate-dependent noise}\label{sec:gate_dependent}

In this section, we broadly justify the validity of
NIST RB in the case of gate-dependent error models.
Since a complete analysis would necessitate 
pages of mathematical developments, we first suggest
the reader to get familiar with
\cite{Wallman2018GD,carignan2018randomized,Merkel2018,proctor2018direct}. 

The essence behind the proof of the decay model 
\cref{eq:nist_decay_model} resides in realizing 
that the RB recovery probabilities evolve as a combination
of decays $\sum_i p_i^m$, where $p_i$ are the
eigenvalues of $\mbb E_{\bf N} G \otimes \tilde{G}~.$
Those eigenvalues are slightly perturbed from
those of $\mbb E_{\bf N} G \otimes G~.$
In the case of the NIST gate-set $\bf N$, the non-zero
eigenvalues are $1,1,1/2$ and $-1/2$. In the 
perturbed case ($\mbb E_{\bf N} G \otimes \tilde{G}$), the first eigenvalues remains $1$, as it translates in the 
trace-preservation property, and the second one becomes
$p_{\bf N} \approx 1$. The two eigenvalues 
which are close to $\pm 1/2$ decay very fast as the
circuit grows, and don't contribute 
to the decay model for $m$ large enough.

The relationship between the decay constant 
$p_{\bf N}$ and the infidelity $r_{\bf N} $ (\cref{eq:nist_infidelity})
can be derived as a straightforward generalization of the analysis derived in \cite{carignan2018randomized}.
Let the eigenvector related with the decay $p_{\bf N}$ be
\begin{align}
    \mbb E_{\bf N} G \otimes \tilde{G} {\rm vec}(L)=p_{\bf N} {\rm vec}(L)~,
\end{align}
where ${\rm vec}(\cdot)$ is the column vectorization. 
In a nutshell, \cref{eq:nist_infidelity}
holds as long as the singular values of $L$ are close to each other, which is shown to be the case in 1-qubit SRB \cite{carignan2018randomized}. The reasoning, which
pertains for the NIST gate-set, goes as 
follow. Let $\Pi_{\rm tr}$ be the $3\times 3$ projector
on the traceless hyperplane (the Bloch space). 
Given the spectrum of $\mbb E_{\bf N} G \otimes \tilde{G}$, we have that:

\begin{align}
    L \approx \mbb E \left(\frac{\tilde{G}_m \tilde{G}_{m-1} \cdots \tilde{G}_1 G_{m:1}^{-1}}{p_{\bf N}^m} \right) \Pi_{\rm tr} ~, \label{eq:L}
\end{align}

for $m$ large enough so the r.h.s converges. Indeed,
performing $p_{\bf N}^{-1}\mbb E_{\bf N} \Pi_{\rm tr} G \otimes \tilde{G}$ multiple times converges very 
quickly to a rank-1 projector onto the desired eigenspace. Since $L$ is the result of a reasonably short sequence of noisy operations ( say $m= -2\log (r) /\log(2)$ ), it is proportional to a high-fidelity channel, for which the singular values are close to each other (at least in the single-qubit case).

\bibliography{qcvv}

\end{document}

%% file: macros.tex
\usepackage{hyperref,graphicx, amsmath, amssymb, amsthm, color, bm, bbm,dsfont}
\usepackage{cleveref}
\usepackage{amsthm}
\usepackage{thmtools}
\usepackage{mdframed}
\definecolor{lichtgrijs}{gray}{0.95}

\declaretheoremstyle[
notefont=\bfseries, notebraces={}{},
bodyfont=\normalfont,
headformat=\NAME~\NUMBER:\NOTE
]{nopar}

\newmdtheoremenv[style=mystyle]{theorem}{Theorem}

\declaretheorem[style=nopar]{protocol}

\newcommand{\mynobreakpar}{\par\nobreak\@afterheading}
\usepackage{float} 

\usepackage{cleveref}
\usepackage{tcolorbox}
\usepackage{enumerate} 
\tcbuselibrary{theorems}
\newtcbtheorem[]{boxproto}{\emph{Protocol}}{colback = blue!10, colframe = blue!50!black, fonttitle = \bfseries}{proto}
\crefname{tcb@cnt@boxproto}{protocol}{protocols}

\theoremstyle{plain}

\theoremstyle{definition}

\crefname{thm}{theorem}{theorems}
\crefname{protocol}{protocol}{protocols}
\crefname{subroutine}{subroutine}{subroutines}

\definecolor{nblue}{rgb}{0.2,0.2,0.7}
\definecolor{ngreen}{rgb}{0.1,0.5,0.1}
\definecolor{nred}{rgb}{0.8,0.2,0.2}
\definecolor{nblack}{rgb}{0,0,0}

\DeclareMathOperator{\tr}{\mathrm{Tr}}
\usepackage{scalerel}
\usepackage{stackengine,wasysym}

\newcommand{\mbb}[1]{\mathbb{#1}}

\newcommand{\hidden}[1]{}